\documentclass[aps,pre,reprint,graphicx]{revtex4-1}
\usepackage{graphicx}
\usepackage[dvipdfm,colorlinks,linkcolor=blue,citecolor=blue,urlcolor=blue]{hyperref}
\draft 
\usepackage{CJKutf8}

\begin{document}
\begin{CJK}{UTF8}{gbsn}


\title{Spectral control of high harmonics from relativistic plasmas using bicircular fields} 



\author{Zi-Yu Chen}
\email[]{ziyuch@caep.ac.cn}
\affiliation{National Key Laboratory of Shock Wave and Detonation Physics, Institute of Fluid Physics, China Academy of Engineering Physics, Mianyang 621999, China}
%


\date{\today}

\begin{abstract}
We introduce two-color counterrotating circularly polarized laser fields as a new way to spectrally control high harmonic generation (HHG) from relativistic plasma mirrors. Through particle-in-cell simulations, we show that only a selected group of harmonic orders can appear owing to the symmetry of the laser fields and the related conservation laws. By adjusting the intensity ratio of the two driving field components, we demonstrate the overall HHG efficiency, the relative intensity of allowed neighboring harmonic orders, and the polarization state of the harmonic source can be tuned. The HHG efficiency of this scheme can be as high as that driven by a linearly polarized laser field.

\end{abstract}

\pacs{}

\maketitle 

\section{Introduction}
High harmonic generation (HHG) from relativistically intense laser irradiating plasma surfaces is an extreme nonlinear process, which converts the fundamental driving laser frequency to its harmonics, which can span thousands of harmonic orders\cite{Teubner2009,Gordienko2004,Gordienko2005,Dromey2007}. One of the main mechanisms is the so called relativistically oscillating mirror (ROM) model\cite{Bulanov1994,Lichters1996,Baeva2006,Pukhov2006,Dromey2006}. The physical picture can be interpreted as relativistic Doppler up-shifting of the laser field reflected by the plasma surface, which acts as a mirror oscillating with speed close to that of light. The harmonics are usually emitted in the form of a comb in the frequency domain and attosecond pulses in the time domain due to the broad spectral range. This thus provides a powerful radiation source in the extreme ultraviolet and x-ray spectral region with attosecond duration. While many an investigation has been carried out to control the temporal structure of the harmonics, e.g., to obtain an isolated single attosecond pulse\cite{Baeva2006b,Rykovanov2008,Liu2008,CL2017}, methods to control the HHG spectrum remain limited. Among the reported spectral control schemes, selected enhancement of HHG has been achieved by modifying the plasma density ramp\cite{Dromey2009} or plasma surface morpha\cite{LD2009,LD2010,Cerchez2013,Fedeli2017,Zhang2017}.

Here we consider controlling the harmonic spectrum by structuring the laser field with a two-color field. Recently, a two-color field scheme, i.e., two paralleled linearly polarized (LP) fields, has been introduced to relativistic plasma HHG\cite{Edwards2014}, which shows great potential to enhance the resulting high harmonic yield\cite{Yeung2017}. Our approach is different, which is based on using circularly polarized (CP) two-color driving pulses rotating in opposite directions. Interestingly, it is known that the ROM HHG process is almost completely suppressed by using one single CP pulse at normal incidence\cite{Yeung2014}, because the laser ponderomotive force in this case contains only slowly varying components following the pulse envelop. The situation may change drastically when combining two CP pulses with different frequency and helicity, as the pulses superposition can lead to very different field patterns as well as charge trajectories. It has already been shown both theoretically\cite{Long1995,Becker1999,Milosevic2000a,Medisauskas2015} and experimentally\cite{Eichmann1995,Fleischer2014,Kfir2015} that the bicircular configuration with one fundamental frequency $\omega$ and its counterrotating second harmonic $2\omega$ can be effective in CP HHG from atomic gases, whereas the HHG process is also greatly suppressed driven by a one-color CP pulse, due to an electron's lateral spreading and reduced probability of re-collision with its parent ion. It is demonstrated that only specific sets of harmonic orders can be displayed and selectively enhanced in the harmonic spectrum driven by such bicircular fields\cite{Eichmann1995,Fleischer2014,Kfir2015,Baykusheva2016,Reich2016,Dorney2017}.
Although such a field configuration has been studied extensively in gas HHG, to our knowledge it has not been addressed for surface plasmas HHG before. It is not clear whether this scheme should work in the relativistic plasma HHG regime or not, since the two HHG mechanisms are fundamentally different. Therefore, the results obtained before within the gas HHG regime cannot be applied directly to relativistic plasma HHG without validation. 

In this work, we use \textit{ab initio} particle-in-cell (PIC) simulations to confirm the effectiveness of such a scheme in the relativistic plasma HHG regime and find that the similar selection rules discovered in gas HHG are also applicable to relativistic plasma HHG. Thus the bicircular fields can be used to spectral control high harmonics from relativistic plasmas. The harmonic spectral features, such as the appeared harmonic orders and their helicity, are found to be governed by the symmetry of the laser fields and the related conservation laws. As a result, selective harmonic enhancement and CP harmonic generation can be achieved. It is also shown such an HHG process can have a high efficiency although using CP pulses and under normal incidence interactions.  

\section{Simulation setup}

\begin{figure}[htbp]
\includegraphics[width=0.35\textwidth
]{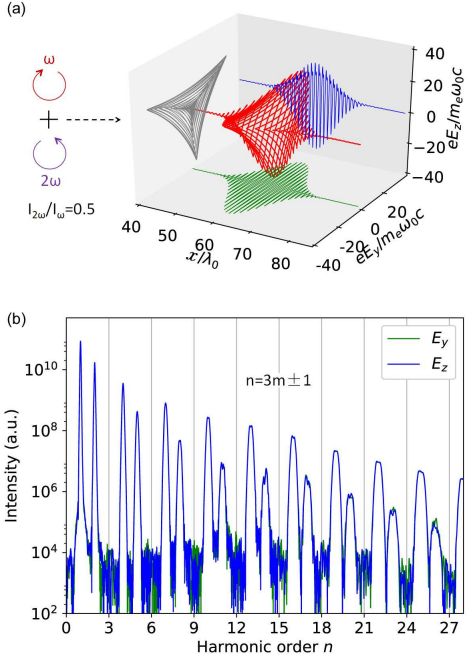}
\caption{\label{spec} (a) Electric field waveform of the incident laser pulses propagating along the $x$ axis, composed of two-color (fundamental frequency and its second harmonic) counterrotating circularly polarized fields. (b) The corresponding high harmonic spectrum from surface plasmas.}
\end{figure}

The simulations were performed using the one-dimensional (1D) PIC code VLPL (Virtual Laser Plasma Lab)\cite{Pukhov1999}. The driving laser fields with fundamental frequency $\omega_0$ (corresponding to wavelength $\lambda_0=$ 800 nm) and its second harmonic $2\omega_0$ are circularly polarized in the same plane but rotate in opposite directions:
\begin{equation}
\textbf{E}(t) = \frac{1}{2i}(E_1 e^{i\omega_0 t} \hat{\textbf{e}}_{-} +E_2 e^{2i\omega_0 t} \hat{\textbf{e}}_{+})+c.c.,
\end{equation}
where $\hat{\textbf{e}}_{\pm}=(\hat{\textbf{e}}_y \pm i \hat{\textbf{e}}_{z}) / \sqrt{2}$ with $\hat{\textbf{e}}_y$ and $\hat{\textbf{e}}_z$ the unit vectors along $y$-
and $z$-axis, respectively. The three-dimensional (3D) electric field waveform is shown in Fig.~\ref{spec}(a), where the intensity ratio of the two driving frequencies $I_{2\omega}/I_{\omega}=0.5$ with $I_{\omega}=2.8\times 10^{21}$ W/cm$^2$ and $I_{2\omega}=1.4\times 10^{21}$ W/cm$^2$. The Lissajous figure projected in the $y-z$ polarization plane (in gray) displays a characteristic feature of threefold spatio-temporal symmetry. The electric field experienced by the plasma electrons acts roughly as three LP pulses rotated by 120$^{\circ}$ three times per optical cycle. This largely modified LP field pattern can give rise to harmonic emission as will be shown later. The temporal intensity profile has a Gaussian envelope with a full-width at half-maximum pulse duration of 30 fs. The laser pulse is normally incident onto a plasma target, of which the peak density is $n_0=200 n_c$ and the thickness is 200 nm. Here $n_c=m_e \omega_0^2 / 4\pi e^2$ is the plasma critical density with respect to the fundamental laser frequency. An exponential density $n_e(x) = n_0 \exp(x/L_s)$ exits in the front surface of the plasma slab, where the preplasma scale length $L_s=0.01\lambda_0$. The ions are taken to be immobile. The simulation cell size is $\lambda_0/1000$, and each cell is filled with 100 macroparticles. An absorption boundary condition is adopted for both fields and particles.

\begin{figure}[htp]
\includegraphics[width=0.49\textwidth
]{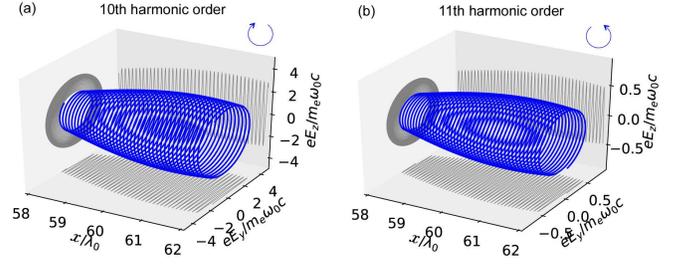}
\caption{\label{10v11} Electric field waveform of the (a) 10th and (b) 11th harmonic shown in Fig.~\ref{spec}(b). In order to see the helicity clearly, only parts of the waveform are shown. The rotation direction is defined as seen by a wave receiver, and the harmonic waves propagate along the -$x$ axis.}
\end{figure}

\section{Results and discussion}
Figure~\ref{spec}(b) shows the harmonic spectra obtained by fast Fourier transform of the reflected electric field. It is clearly seen that only the harmonic orders with $n=3m\pm 1$ ($m=1,2,3,\cdots$) have been generated, while every third one with $n=3m$ is missing. As the plasma is isotropic and the interaction is at normal incidence, this unique harmonic feature can be explained by the symmetry of the driving field as follows. We follow the arguments of Kfir \textit{et al}\cite{Kfir2015} and Pisanty \textit{et al}\cite{Pisanty2014}. The laser field vector, rotated by 120$^{\circ}$ per cycle in the polarization plane, satisfies,
\begin{equation}
\textbf{E}^{\text{L}}(t+T_0/3) = \hat{R}_{(2\pi/3)} \textbf{E}^{\text{L}}(t),
\end{equation}
where $\hat{R}_{(2\pi/3)}$ is the $120^{\circ}$ rotation operator.  
Assuming the emitted harmonic field conforms to the same dynamical symmetry and by taking Fourier transform, we can obtain the spectral field of the $n$th-order harmonic, $\textbf{E}^{\text{H}_{\omega}}_n$, satisfying the following eigenvalue equation
\begin{equation}\label{eigen}
e^{-2\pi i n/3} \textbf{E}^{\text{H}_{\omega}}_n  = \hat{R}_{(2\pi/3)} \textbf{E}^{\text{H}_{\omega}}_n.
\end{equation}
The solutions are (1) $n=3m+1$, so that $e^{-2\pi i n/3} \textbf{E}^{\text{H}_{\omega}}_{n}= e^{-2\pi i/3} \textbf{E}^{\text{H}_{\omega}}_{n}$ and (2) $n=3m-1$, so that $e^{-2\pi i n/3} \textbf{E}^{\text{H}_{\omega}}_{n}= e^{+2\pi i/3} \textbf{E}^{\text{H}_{\omega}}_{n}$.
Apart from these two sets of harmonic orders, the harmonic orders with $n=3m$ do not satisfy Eq.~(\ref{eigen}) and are therefore forbidden by the threefold symmetry. 

\begin{figure*}[htbp]
\includegraphics[width=0.9\textwidth
]{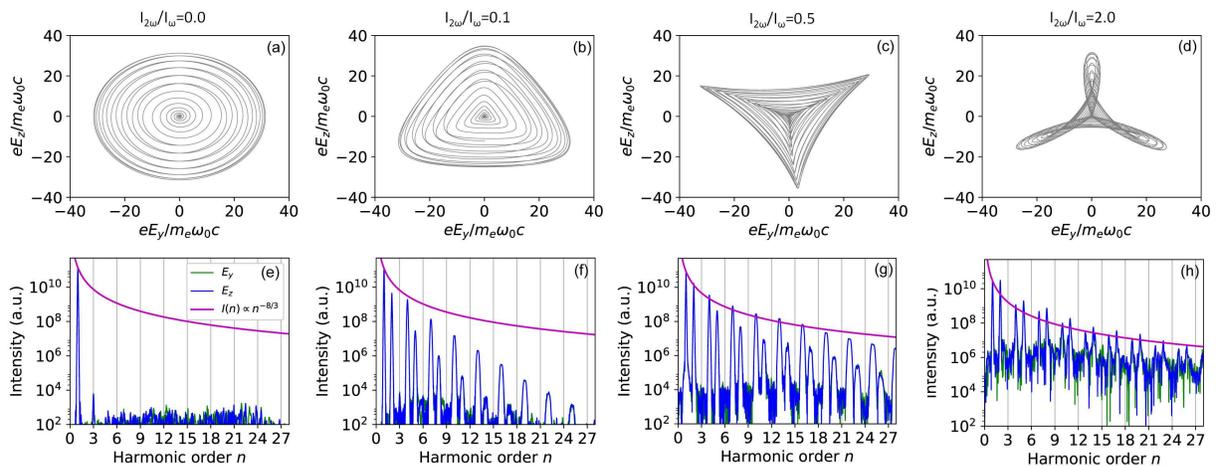}
\caption{\label{lsj} Polarization plane projections of the laser electric fields generated with intensity ratios (a) $I_{2\omega}/I_{\omega}=0.0$, (b) $I_{2\omega}/I_{\omega}=0.1$, (c) $I_{2\omega}/I_{\omega}=0.5$, and (d) $I_{2\omega}/I_{\omega}=2.0$. The total intensity is kept constant ($I_{\omega}+I_{2\omega}=4.2\times 10^{21}$ W/cm$^2$).
(e)-(h) The generated harmonic spectra corresponding to the lasers of panels (a)-(d).}
\end{figure*}

The preceding selection rules are also consistent with the conservation laws for energy, parity and spin angular momentum\cite{Fleischer2014,Pisanty2014}. Considering the HHG process in a simple photon-exchange picture, conservation of energy implies each harmonic frequency can be expressed in the form of
\begin{equation}
\Omega^{\text{H}}_{(n_1,n_2)} = n_1 \cdot \omega_0 + n_2 \cdot 2 \omega_0 ,
\end{equation} 
where $n_1$ and $n_2$ are integers. Applying parity operation on both the incoming and outgoing photons, we have
\begin{equation}
P [h_{(\Omega^{\text{H}}_{(n_1,n_2)})}] = -1;
\end{equation}
\begin{equation}
P [h_{(n_1 \cdot \omega_0 ; n_2 \cdot 2 \omega_0)}] = (-1)^{n_1+n_2},
\end{equation}
where $P$ is parity and $h$ represents helicity. Conservation of parity requires that $n_1 + n_2$ must be odd. In addition, conservation of spin angular momentum leads to
\begin{equation}
\sigma^{\text{H}}_{(n_1,n_2)} = n_1 \sigma_1 + n_2 \sigma_2.
\end{equation}
Since the $\omega_0$ and $2\omega_0$ driving pulses are counterrotating, we can set $\sigma_1 = 1, \sigma_2 = -1$. For photon spin, $ -1 \leq \sigma_{(n_1,n_2)} \leq 1 $. Then it is easily seen that $n_2 -1 \leq n_1 \leq n_2 +1$, which together with the parity requirement means that $n_1$ and $n_2$ must differ by unity. Therefore, only two cases are allowed: (1) $n_1=m+1$ and $n_2=m$, leading to $(m+1)\omega_0 + m \cdot 2\omega_0 = (3m+1)\omega_0$ and (2) $n_1=m$ and $n_2=m+1$, leading to $m\omega_0 + (m+1)\cdot 2\omega_0 = (3m+2)\omega_0$. Note $3m+2$ is equivalent to $3m-1$. These two selection rules also indicate that the $n=3m+1$ harmonic orders have the same helicity with the fundamental field, while the $n=3m+2$ orders co-rotate with the $2\omega_0$ field. These are consistent with the eigenvalues $e^{-2\pi i/3}$ and $e^{+2\pi i/3}$ of Eqs.~(\ref{eigen}), which also imply the $n=3m+1$ and $n=3m-1$ harmonics have opposite helicity.

The aforementioned harmonic helicity is confirmed by the simulation results. Figure~\ref{10v11} shows the 3D electric field waveform of a representative group of neighboring harmonics ($n=10$ and 11). To see the helicity clearly, only parts of the waveform are shown. Each harmonic is circularly polarized. The helicity of the 10th harmonic is the same as the fundamental field, while the 11th harmonic is the same as the second harmonic field. The other harmonic groups share the same features. We note although the adjacent harmonics have alternating helicity, the relative intensity of neighboring harmonics is not equal (see Fig.~\ref{spec}(b)). Here the $n=3m+2$ harmonics are weaker than the $n=3m+1$ harmonics, due to the $2\omega_0$ field being less intense than the $\omega_0$ field. As a result, harmonics of all the groups combined can be circularly or elliptically polarized. Otherwise, the combined harmonic fields should be close to linear polarization\cite{Milosevic2000}. This offers an alternative way to generate CP HHG and attosecond pulses from relativistic plasmas\cite{Chen2016,Ma2016,Chen2018}. Such radiation sources can find important applications in ultrafast measurement of chiral molecules\cite{Cireasa2015} and magnetic materials\cite{Willems2015}. In addition, it was believed previously that CP HHG from relativistic plasmas could be achieved only under oblique incidence interactions\cite{Chen2016}. The results here show it is also possible to generate CP surface harmonics at normal incidence. 

Next, we show that the harmonic intensity can be modulated by adjusting the intensity ratio of the two driving pulses. We keep the total intensity of the two fields constant ($I_{\omega}+I_{2\omega}=4.2\times 10^{21}$ W/cm$^2$) as the ratio $I_{2\omega}/I_{\omega}$ varies from 0 to 0.1, 0.5, and 2. The remaining parameters are the same as in Fig.~\ref{spec}. The Lissajous figures of the superposed electric fields are displayed in Figs.~\ref{lsj}(a)-(d). The corresponding harmonic spectra are shown in Figs.~\ref{lsj}(e)-(h). For the case of $I_{2\omega}/I_{\omega}=0$, i.e., one-color CP driving laser only, high harmonics with orders $n>3$ are indeed strongly suppressed, almost at the noise level. As the laser intensity ratio is increased, the harmonic intensity is significantly enhanced. For $I_{2\omega}/I_{\omega}=2.0$, the spectral intensity of all the allowed harmonic orders is close to the $I(n) \propto n^{-8/3}$ scaling, which is theoretically derived for the case of LP driving lasers\cite{Baeva2006} and already confirmed by experimental results\cite{Dromey2006}. This indicates that the scheme of bicircular driving lasers can also be very efficient in high harmonic generation. The enhancement of HHG efficiency can be attributed to the different field vector pattern induced by a different laser intensity ratio. As the laser field experienced by the plasma electrons gets closer to linearly polarized, stronger harmonic strength can be expected. Also, with increasing $I_{2\omega}$, the relative intensity of allowed neighboring harmonic orders $I_{3m+2}/I_{3m+1}$ also increases. This would change the overall polarization state of the harmonic source.

\begin{figure}[htbp]
\includegraphics[width=0.35\textwidth
]{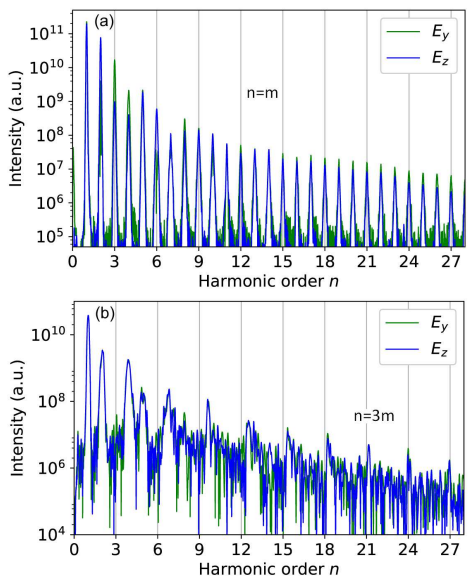}
\caption{\label{3m} High harmonic spectrum for the case of (a) oblique incidence with incidence angle $\theta = 45^{\circ}$ and (b) plasma scale length of $L_s=0.15\lambda_0$. The other simulation parameters are the same as in Fig.~\ref{spec}.}
\end{figure}

Although the Lissajous figures in Figs.~\ref{lsj}(b)-(d) have a very different appearance, they all keep a threefold symmetry. Therefore, the harmonic spectra in Figs.~\ref{lsj}(f)-(h) have the same selection rule as $n=3m\pm 1$.
When the threefold symmetry is lost, the selection rule described above no longer holds. A variety of harmonic channels can open up. The simplest way to break the threefold rotational symmetry of the interaction is to change the incidence geometry from normal to oblique. In the oblique incidence case, the symmetry breaking can be seen more clearly by considering the interaction in the 1D case. Using Bourdier's method to take Lorentz transformations from the laboratory frame to a moving frame\cite{Bourdier1983}, the laser pulse can be transformed to be at normal incidence (wave vector $\textbf{k} = k \hat{\textbf{e}}_x$). At the same time, the plasmas initially stream along the $-\hat{\textbf{e}}_y$ direction while they are stationary along the $\hat{\textbf{e}}_z$ direction. Thus the rotational symmetry in the ($y-z$) polarization plane is broken. Consequently, the previously symmetry-forbidden $n=3m$ orders can appear. Figure~\ref{3m}(a) shows the bicircular field-driven harmonic spectrum for the case of oblique incidence with incidence angle $\theta = 45^{\circ}$. The other simulation parameters are the same as in Fig.~\ref{spec}. We see the harmonic orders with $n=3m$ are indeed shown in this case.

It is interesting to note only the $n=3m$ harmonic orders appear in the spectrum for $n>15$ shown in Fig.~\ref{3m}(b), where the plasma scale length $L_s=0.15$ while the other simulation parameters are the same as in Fig.~\ref{spec}. The physical origin for this "abnormal" phenomena, however, is not from symmetry-based selection rules, but more likely from plasma effects. On the one hand, the appearance of $n=3m+1$ and $n=3m+2$ orders and the absence of $n=3m$ orders for $n<10$ reflect the preservation of the threefold symmetry. On the other hand, as $L_s$ becomes longer, significant plasma waves can be excited and are able to couple to the incident wave. Consequently, various parametric instabilities and self-phase modulation effects\cite{Dollar2013}, together with the chirp effect due to the motion of surface plasmas\cite{Behmke2011}, can result in harmonic spectral modulation, splitting, and frequency shifting. The gradual frequency shifting from the $n=3m+1$ orders to match the $n=3m$ orders is evident when the harmonics of $n>12$ are compared. The absence of the $n=3m+2$ orders in this case may be explained as too weak to distinguish. For even longer $L_s$, however, the whole high harmonic structure would disappear due to strong plasma modulation effects and low efficiency of the ROM mechanism in this case.

\section{Conclusions}
In summary, we introduce bicircular counter-rotating two-color fields to HHG in the relativistic plasma regime for the first time. We numerically demonstrate, by adjusting the intensity ratio of the two driving fields, that the properties of the high harmonics, including the allowed harmonic orders and their overall intensity, the relative intensity of allowed neighboring harmonic orders, as well as the polarization state of the harmonics, can be controlled. The underlying mechanisms can be understood as a result of the symmetry effect and related conservation laws. This work thus provides an alternative way to spectral control surface harmonics. Also, it offers a way to generate CP high harmonics and/or attosecond pulses from relativistic plasmas. Moreover, it also shows it is possible to generate CP harmonics from relativistic plasmas at normal incidence with high efficiency. This scheme opens up more possibilities in controlling the surface harmonics, as more degrees of freedom can be adjusted, such as the frequency, ellipticity, and relative phase of the two drivers in the counterrotating two-color fields.

\section*{acknowledgments}
This work was supported in part by the National Natural Science Foundation of China (Grants No. 11505172 and 11705185) and the Presidential Fund of China Academy of Engineering Physics (Grant No. YZJJLX2017002).


\end{CJK}
\end{document}